# Two amateur astronomers at Berkeley

Amelia Carolina Sparavigna
Department of Applied Science and Technology
Politecnico di Torino

The book on Mechanics of the Physics at Berkeley, by C. Kittel, W.D. Knight and M.A. Ruderman [1], is proposing at the end of its first chapter some problems of simple astronomy within the solar system. The discussion begins with two amateur astronomers who set for themselves the goal of determining the diameter and mass of the Sun. Here we discuss the problems proposed by the book and some other matters on ancient and modern astronomical studies of the solar system.
The starting point for the two astronomers to find the dimensions of the solar system is the measure of some fundamental quantities. For the time intervals, we have the periods of rotation and revolution of the Earth, and the period of revolution of the Moon. For the distances, as told in [1], the astronomers soon realize that, to begin with, it is necessary to know the Earth's radius.

**The radius of Earth**
The two amateur astronomers are on the same geographical meridian and know precisely their position. Their distance $D$ on the meridian is 804 km. They communicate using wireless-sets. The southern observer, $O_S$, chooses a star that at a certain instant of time is passing at the Zenith. The observer sees the star with the direction coincident with the Z-axis At the same time, the northern observer, $O_N$, sees the same star at a certain angle from his Zenith. In Figure 1, the $O_N$ Zenith is the X-axis. The light coming from the star has the direction Y. This direction is parallel to Z, because we imagine the star at a so large distance that its light rays can be considered as parallel.
Supposing a measured angle of 7.2 degrees, the radius $R$ of the Earth turns out to be $6.4 \times 10^8$ cm. In fact, $D = R \Theta$ and then $R = D / \Theta$.

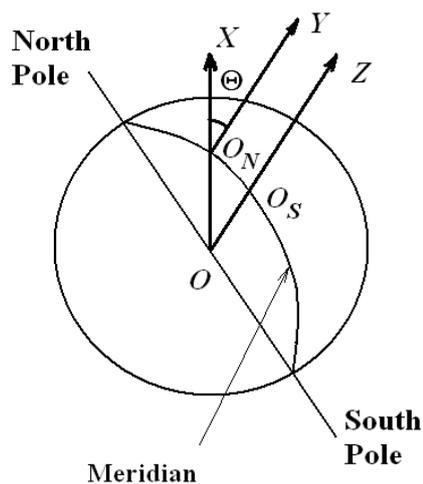

Fig.1

This is, more or less, what Eratosthenes did to evaluate the circumference of the Earth. Eratosthenes of Cyrene, Libya, (c.276 BC – c.195 BC, [2]), Greek mathematician, poet, and astronomer, was the third chief librarian of the Great Library of Alexandria of Egypt. Eratosthenes was the first to study the discipline of geography as we consider it nowadays [3]. He proposed a system of latitude and longitude, evaluated the tilt of the Earth's axis, and, as previously told, measured the radius of the Earth.

Eratosthenes knew that on the summer solstice at local noon in the Egyptian city of Syene, the modern Aswan on the Tropic of Cancer, the sun would be at the Zenith. That is, there is no shadow cast by the gnomon of a sun dial, at noon on the day of the solstice. He also knew that in Alexandria, at noon of the solstice, the sun was at an angular position of 1/50th of a circle (7.2 degrees) South of the Zenith [4]. To evaluate this angle $\Theta$ he observed the shadow of a gnomon (someone imagines an obelisk). He probably measured two lengths: the length $L$ of a gnomon perpendicular to the ground as a plumb and the length of its shadow $l$. From the ratio $l/L$, he had $tan\ \Theta$, that, for an angle of 7.2 degrees, is equal to 0.1263. In radians, the angle $\Theta$ is equal to 0.1256. The difference between the angle and its tangent is then 5/1000. It is possible to use $tan\ \Theta$ as the value of $\Theta$. Probably, Eratosthenes' measures were not more precise.

Assuming that the Earth was spherical and that Alexandria ($O_N$) was northern of Syene ($O_S$) on the same meridian, he deduced the arc distance from Alexandria to Syene be 1/50th of the total circumference of the Earth. For a distance between the cities of 5000 stadia (about 804 km), we obtain the value of $6.4 \times 10^8$ cm for the radius of the Earth.

Information on the Eratosthenes' method is coming from a chapter on the winds of "The Architecture", by Marcus Vitruvius Pollio [5]. He is writing that "Eratosthenes of Cyrene, employing mathematical theories and geometrical methods, discovered from the course of the sun, the shadows cast by an equinoctial gnomon, and the inclination of the heaven that the circumference of the earth is two hundred and fifty-two thousand stadia". Vitruvius is referring of "equinoctial gnomon": did Eratosthenes the measures also on equinoxes?

Pliny the Elder in his "Natural History" [6], reported about Eratosthenes too. In a chapter entitled "when and where there are no shadows", Pliny tells that "in the town of Syene, which is 5000 stadia south of Alexandria, there is no shadow at noon, on the day of the solstice; and that a well, which was sunk for the purpose of the experiment, is illuminated by the sun in every part. Hence it appears that the sun, in this place, is vertical". Probably Pliny read about the Eratosthenes' measures, but he did not discuss them. He just wrote about the shadows. In another chapter, entitled "The dimensions of the Earth", Pliny reported the value of the radius of the Earth. Pliny tells that "Eratosthenes, a man who was peculiarly well skilled in all the more subtle parts of learning, and in this above everything else, and a person whom I perceive to be approved by every one, has stated the whole of this circuit to be 252,000 stadia, which, according to the Roman estimate, makes 31,500 miles. The attempt is presumptuous, but it is supported by such subtle arguments that we cannot refuse our assent."

Who discussed the Eratosthenes' method was Cleomedes. He was a Greek astronomer, author of the book "On the Circular Motions of the Celestial Bodies", a basic astronomy textbook in two volumes [7]. Historians have suggested that he lived between the 1st century BC and 400 AD. Cleomedes discussed the lunar eclipses, concluding that the shadow on the Moon suggests a spherical Earth. In this book we find the Eratosthenes' measure of the Earth's circumference. If Cleomedes lived during the 1st century BC, his book was probably the source of the Vitruvius' and Pliny's discussions. Cleomedes is in fact reporting in detail the method with the sundials [8].

According to Eratosthenes, the Earth has a circumference of 252,000 stadia. The problem is then to know the size of the "stadion" that he used. The common Attic stadion was about 185 meters [9], giving a circumference of 46,620 km. Assuming that Eratosthenes used the Egyptian stadion [10] of about 157.5 m, his measurement gives 39,690 km.

**The speed of the Moon**

The second quantity the two astronomers at Berkeley want to measure is the speed of the Moon during its orbit about the Earth. To determine this speed, the two observers at two different places on the Earth, $O_O$ and $O_E$, measure the time of the occultation of a star by the Moon.

An occultation in astronomy occurs when a celestial body is hidden by another body passing between it and the observer. In particular, an occultation happens when an apparently larger body passes in front of an apparently smaller one. On the other case, we have a transit when an apparently smaller body passes in front of an apparently larger one.

The term occultation is used in particular to describe the events when the Moon passes in front of a star. Since the Moon has no atmosphere, a star that is occulted by the Moon will disappear or reappear in 0.1 seconds or less on the Moon's edge, that is called "limb". Moreover, those occultations that take place on the Moon's dark limb render the event easy to measure, due to the lack of glare. The orbit of the Moon is inclined to the ecliptic. For this reason, stars with an ecliptic latitude of less than about 6.5 degrees may be occulted. There are three first magnitude stars that can be occulted: they are Antares, Regulus and Spica. Aldebaran too can be occulted [11].

Let us go back to the problem of evaluating the speed of the Moon. To avoid some complexities of calculus, the astronomers assume that the Moon and the star are lying on the ecliptic. This is the orbital plane of the Earth. Moreover, the two astronomers observe the star occultation at the midnight of full Moon. The star passes exactly on a line corresponding to the diameter of the Moon. The curvature of the Earth and the atmospheric refraction effects are neglected.

The geometry of the observation is shown in Fig.2.

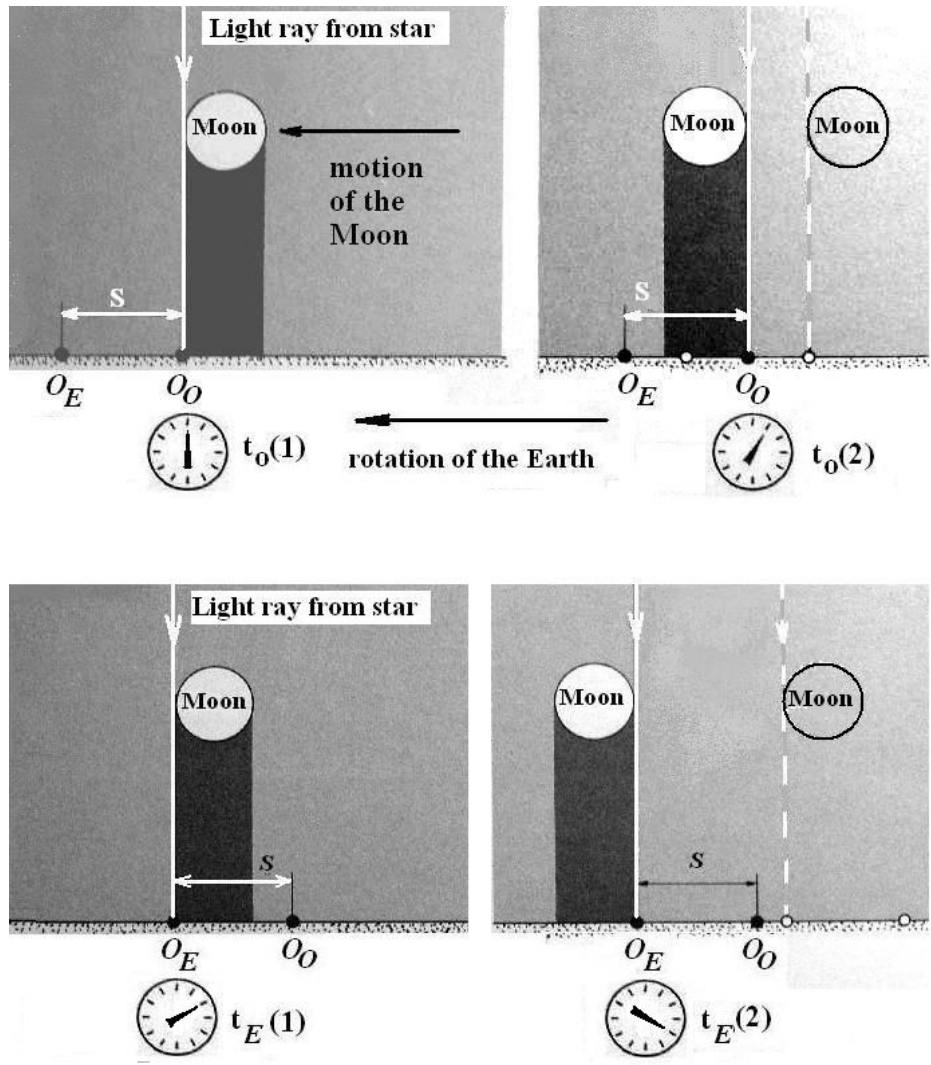

Fig.2 At the time $t_O(1)$ the star is occulted by the Moon for the western observer $O_O$. The star reappears at the time $t_O(2)$. At time $t_E(1)$, the occultation starts for the eastern astronomer $O_E$. The occultation finishes at the time $t_E(2)$.

In this geometry the star is at a so large distance from the Earth that the light rays from the star are parallel. The two astronomers are $O_O$ at a western position and $O_E$ at a eastern position. Each observers has a clock. The clocks are synchronized. $O_O$ measures two times: $t_O(1)$ when the star is occulted by the Moon and $t_O(2)$ when the star reappears. Observer $O_E$ does the same, obtaining $t_E(1)$ and $t_E(2)$. Book [1] tells that they determine the speed of the Moon as:

$$v_{Moon} = v_o + \frac{S}{t_E(1) - t_o(1)} \tag{1}$$

$S$ is the distance $O_O O_E$. But the two observers have their own speed $v_o$, because the Earth is rotating. Equation (1) is the equation of the relative motion. In fact:

$$v_{Moon} = v_o + v' \tag{2}$$

where it is assumed for simplicity a translational relative motion to avoid vectors. In this equation, $v'$ is the speed of the Moon measured by the two astronomers, which is defined as the distance travelled $S$ divided by the time interval $(t_E(1) - t_O(1))$. Since we know the distance and we measure the times, this quantity is known.

We need to calculate $v_o$. The problem of [1] is assuming $O_O, O_E$ both at 30 degrees of latitude North. The Earth is rotating in 24 hours: then the period T is equal to 86400 seconds.

The radius of the Earth is $6.4 \times 10^8$ cm. At the Equator, the speed of an observer on the Earth surface is (Fig.3):

$$v_{sup} = \frac{2\pi}{T} R = \omega R = 0.727 \times 10^{-4}\ sec^{-1}\ 6.4 \times 10^8\ cm = 4.656 \times 10^4\ \frac{cm}{sec} \tag{3}$$

At 30° of latitude North, the radius of the rotation is smaller:

$$R' = R \cos 30° = \frac{\sqrt{3}}{2} R \tag{4}$$

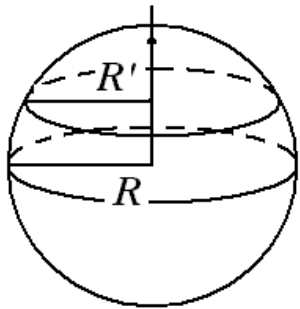

Fig.3

The speed is therefore:

$$v_o = \omega R' = 0.727 \times 10^{-4}\ sec^{-1}\ \frac{\sqrt{3}}{2} 6.4 \times 10^8\ cm = 4.04 \times 10^4\ \frac{cm}{sec} \tag{5}$$

From the table given in the book [1], we have that: $t_O(1) = 0.0$ *min*, $t_O(2) = 95.6$ *min*, $t_E(1) = 22.0$ *min*, and $t_E(2) = 117.7$ *min*. Let us assume again a distance $S = 804 \times 10^5$ *cm*. So the speed of the Moon is:

$$v_{Moon} = v_o + \frac{S}{t_E(1) - t_o(1)} = 4.04 \times 10^4 \frac{cm}{sec} + \frac{804 \times 10^5 \, cm}{22 \times 60 \, sec} = 10.1 \times 10^4 \frac{cm}{sec} \qquad (6)$$

This is the orbit speed of the Moon in the frame moving with the centre of the Earth.
Another question of the book [1] is to find the diameter of the Moon. We have the speed $v'$ of the Moon in the rotating Earth frame. During the time interval $t(2) - t(1)$, the Moon moves a distance equal to its diameter. Therefore, the diameter is given by the following equation:

$$2R_{Moon} = v' \cdot (t_o(2) - t_o(1)) = 3.48 \times 10^8 \, cm \qquad (7)$$

**The distance of the Moon**
This is another question of the Berkeley book find the distance of the Moon. First of all, we need to observe the period of its revolution. In fact we have its speed and then, assuming a circular orbit of the Moon, we can calculate the orbital radius.
The period of revolution is approximately of 27.3 days. The period is then:

$$T_{Moon} = 27.3 \times 24 \times 60 \times 60 \, sec = 2.36 \times 10^6 \, sec \qquad (8)$$

Assuming a circular orbit, we call $D$ the orbit radius, which is therefore the distance of the Moon, The orbital speed is given by:

$$v_{Moon} = \omega D = \frac{2\pi}{T} D \qquad (9)$$

We have already measured this speed and then:

$$D_{Moon} = \frac{T}{2\pi} v_{Moon} = \frac{2.36 \times 10^6 \, sec}{2\pi} 10.1 \times 10^4 \frac{cm}{sec} = 3.84 \times 10^{10} \, cm \qquad (10)$$

Let us note that the Moon is moving on an elliptic orbit having an eccentricity equal to 0.0549 [11].

**The Moon's Parallax**
Let us go back to the years 1751-52. We find two astronomers that provided an excellent measure of the distance Earth-Moon. They were the two French astronomers Joseph Jérôme Le François de Lalande (1732-1807) and the Abbé Nicolas Louis de La Caille (1713-1762). They used a triangulation method to determine the distance.
From their measures it is possible to determine the parallax of the Moon too. The lunar parallax is defined as the angle subtended at the distance of the Moon by the radius of the Earth, equal to the angle p in the diagram of Fig.4, left panel. The determination of the Moon parallax was not simple for the two astronomers: they had to measure the angle Θ of a telescope aimed at a specific place on the Moon, the centre of the Crater Copernicus for instance, from the vertical direction (Zenith). And they had to measure their respective angles at two different distant places on the same meridian. Moreover, these measures must be done at the same time, when the Moon is passing on the meridian. In the Figure 4, right panel, we see the two angles $\Theta_1$ and $\Theta_2$ the astronomers had to determine.

Moreover, they had to know the latitude of their positions, $\Phi_1$ and $\Phi_2$ (see Fig.5).

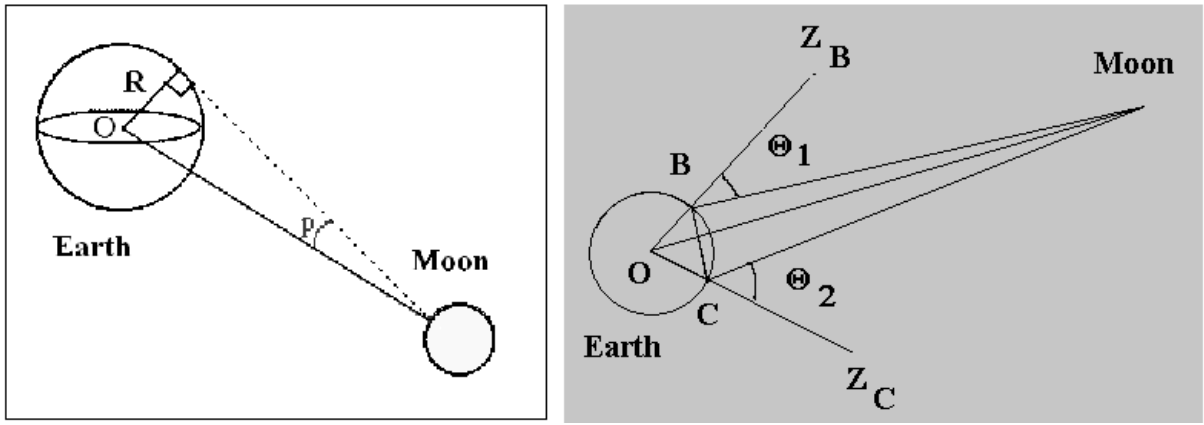

Fig.4 - The left panel shows the Moon parallax p. The right panel shows the geometry of the triangulation method used for determining the distance to the Moon.

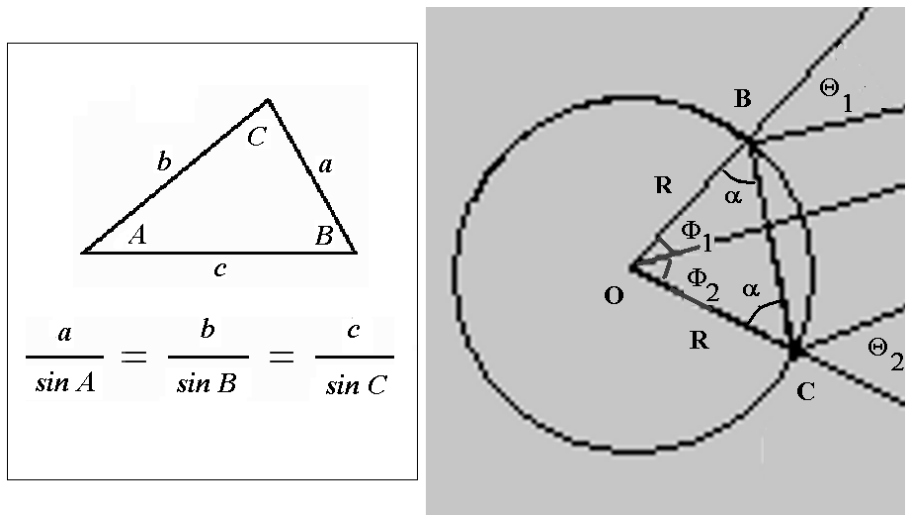

Fig.5 - The left panel shows the sinus rule of a triangle. The right panel shows the angles used in the calculation.

Lalande and La Caille travelled to two observing sites which were located, more or less, on the same meridian. Lalande was at Berlin (latitude $\Phi_1$=52.52° N; longitude 13.24° E), La Caille at the Cape of Good Hope (latitude $\Phi_2$=34.35° S; longitude 18.22° E). They measured several times the angles $\Theta_1$ and $\Theta_2$. In what follows, let us consider the small difference in geographical longitude as negligible and assume therefore that *B* and *C* are located nearly on the same meridian.

The angles measured by Lalande and La Caille, the angles Θ between the direction towards the Moon's centre and the zenith direction is called the "zenith distance" of the Moon.

Let us call $\gamma_1$ the size of the angle OMB and $\gamma_2$ the size of the angle OMC in Fig.4. Let us make a simplified calculation. We suppose that the Earth is spherical. On August 31, 1752, Lalande obtained: $\Theta_1 = 33.11°$ in Berlin; and La Caille: $\Theta_2 = 55.14°$ at the Cape of Good Hope [12].

We have two angles, which are equal as we can see in the Figure 5, named α (BOC is an isosceles triangle). We can evaluate α: the triangle BOC has the sum of the angles $\pi = (\Phi_1 + \Phi_2) + 2\alpha$. From the trigonometry we have that: $\gamma_1 + \gamma_2 = \pi - [\pi - \Theta_1 - \alpha] - [\pi - \Theta_2 - \alpha] = \Theta_1 + \Theta_2 - (\Phi_1 + \Phi_2) =$ (33.11° + 55.14°) – (52.52° + 34.35°) = 88.25° – 86.87° = 1.38 degrees.

In the treatise by Lalande, he obtained the distance from the Earth to the Moon *D*, applying the sinus rule of triangles, as in the following relations:

$$D = R \frac{sin\,\Theta_1}{sin\,\gamma_1} = R \frac{sin\,\Theta_2}{sin\,\gamma_2} = R \frac{sin\,\Theta_1 + sin\,\Theta_2}{sin\,\gamma_1 + sin\,\gamma_2} \qquad (11)$$

Let us note that in the last part of equation, we used the fact that:

$$\frac{a}{b} = \frac{xa}{xb} = \frac{a + xa}{b + xb} = \frac{a}{b}\frac{1+x}{1+x} \qquad (12)$$

Since $\gamma_1$ and $\gamma_2$ are quite small, we can consider the angle instead of the sine of the angle, remembering that in this manner, we have to use the angle in radians:

$$D = R \frac{sin\,\Theta_1 + sin\,\Theta_2}{\gamma_1 + \gamma_2} = R \frac{sin\,33.11° + sin\,55.14°}{\frac{2\pi}{360}1.38} = R \frac{1.367}{0.024} = 3.65 \times 10^{10}\,cm \qquad (13)$$

As evaluated with the approximation of a spherical Earth and neglecting the refraction effects of atmosphere, the radius of the Moon's orbit is $365 \times 10^8$ *cm*. Having this value and knowing the radius of Earth we can easily calculate the parallax p as in the Figure 4, left panel.

Let us note that the value of the distance that we have obtained is between the Moon perigee (approximately $363 \times 10^8$ *cm*) and its apogee ($406 \times 10^8$ *cm*).

**More on the Moon parallax**

There is another manner of determining the Moon parallax. Before the discussion, let me define an interesting astronomical quantity. It is the angular diameter [13], the apparent size of an object as seen from a given position. The sizes of the objects seen in the sky are often given in terms of their angular diameter as seen from Earth, rather than their actual sizes. For instance, for the Sun seen from the Earth, we have andangle ranging from 31.6' to 32.7', and for the Moon, the angle is ranging from 29.3' to 34.1'.

Now, let us determine the lunar parallax by using a lunar eclipse, observing the shadow of the Earth on the Moon. This was the procedure firstly used by Aristarchus of Samos [14]. We will discuss about Aristarchus and his theories of the solar system in a following section.

Reference [15] gives a very clear explanation on this method used by the ancient Greeks to measure the distance between Earth and Moon. First of all, Ref.15 is proposing the following example about an eclipse. Let us imagine to hold up a quarter of dollar, which is one inch in diameter approximately, at the distance where it just blocks out the full Moon's rays from our eye. Experimentally, we find that the right distance is about 108 inches away. The full Moon has the same apparent size as the sun. Therefore, the sunlight is just blocked by a quarter in the same manner. And now, let us imagine to be out in the space, and see the Earth and its shadow (Fig.6). It is a cone, the point of which is the furthest position where the Earth can block all the sunlight. The ratio of the distance of that point to the Earth's diameter is given by the angular size of the Sun. The aperture of the cone is then equal to the angular diameter of the Sun.

Using the angle of 31.6', we have that the cone is 108 Earth diameters long.

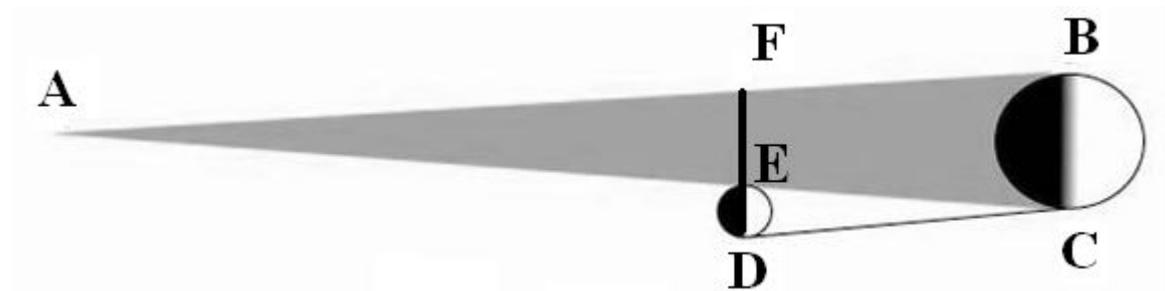

Fig.6

During a total lunar eclipse, the Moon moves passing in the Earth's shadow. By observing the Moon during the eclipse, the ancient Greeks found that the diameter of the Earth's shadow was about two-and-a-half times the Moon's own diameter, at the distance of the Moon. The Greeks used this observation and the following procedure. They knew the fact that the Moon and the Sun have the same apparent size. This means (Fig.6) that the angle ECD is equal to the angle EAF [15].

The observation of the eclipse of the Moon tells that the ratio of FE to ED is 2.5 to 1. We have two similar isosceles triangles FAE and DCE. Therefore we have that AE is 2.5 times as long as EC, and then AC is 3.5 times as long as EC. As previously discussed, AC is 108 Earth diameters in length. This means that EC is 108/3.5 Earth diameters. We have $EC = 395 \times 10^8$ cm.

We see then that the ancient Greek astronomers devised a quite good method, based on the observation of the Earth's shadow on the Moon, to find its distance.

A modern method to determine the parallax is based on two pictures of the Moon [16], taken at exactly the same time from two different locations on Earth. The position of the Moon relative to the stars turns out to be different in the two pictures. Using the orientation of the Earth, these two position measurements and the distance between the two locations on the Earth, the distance to the Moon can be triangulated.

**The Gravitation**

Let us go back to the problems proposed by book [1]. From the astronomical measurements, we have obtained the orbital speed of the Moon during its revolution, its distance from the Earth and of course, its period of revolution. From the gravity law, we know that:

$$F = G \frac{M_{Moon} M_{Earth}}{D^2} \quad (14)$$

where $G$ is the universal constant of gravitation. $M_{Moon}$ and $M_{Earth}$ are the masses of Moon and Earth and $D$ their distance.

We know also that, for a mass $m$ near the surface of the Earth, the acceleration when it is freely falling is:

$$mg = G \frac{mM_{Earth}}{R^2} \rightarrow GM_{Earth} = R^2 g \quad (15)$$

The product $GM_{Earth}$ turns out to be $4.014 \times 10^{20}$ $cm^3\ sec^{-2}$, assuming $g = 980$ $cm\ sec^{-2}$.
The quantity $GM_{Earth}$ is called the standard gravitational parameter of the Earth.
In the case that the Moon is moving in a circular motion, the gravitation is the centripetal force:

$$F = M_{Moon} \frac{v_{Moon}^2}{D} = G \frac{M_{Moon} M_{Earth}}{D^2} \quad (16)$$

From this equation, and from the already evaluated quantities:

$$v_{Moon} = 10.1 \times 10^4 \frac{cm}{sec} \quad (17)$$

$$D = 3.84 \times 10^{10} cm \quad (18)$$

$$v_{Moon}^2 D = GM_{Earth} = \left(10.1 \times 10^4 \frac{cm}{sec}\right)^2 3.84 \times 10^{10} cm = 3.92 \times 10^{20} cm^3 sec^{-2} \quad (19)$$

Besides using the acceleration of gravity, that we can have obtained by measuring the small oscillation of a pendulum, we can have the gravitational product $GM_{Earth}$ from the astronomical observations.
As the book [1] is discussing, to have the value of $G$ one needs another experiment, using a torsion balance, to measure the gravitational force. Reference [1] continues telling that, instead of measuring $G$, the two amateur astronomers decide to estimate the Earth density: from the observation of the surface of the planet, they obtain a density $\rho = 5$ $gm/cm^3$.
The gravitational constant $G$ appeared in Newton's formulation of the law of the universal gravitation. Henry Cavendish measured this constant in 1798, that is, seventy-one years after Newton's death. Cavendish measured $G$, using the torsion balance invented by the geologist John Michell. As reported in [17], he was searching, as our two amateur astronomers, for the average density of the Earth. He measured it as $5.448 \pm 0.033$ times that of water. The density that Cavendish calculated implies a value for the constant $G$ of $6.754 \times 10^{-8}$ $cm^3\ gm^{-1}\ sec^{-2}$.
$G$ is difficult to measure, being the gravitational interaction much weaker than other fundamental forces. Moreover, the experimental apparatus cannot be separated from the gravitational influence of other masses (see the discussion at Ref.17).
Book [1] asks to calculate the mass of the Earth. Assuming a density ρ= 5 *gm/cm³*:

$$M_{Earth} = \rho \frac{4\pi}{3} R^3 = 5 \frac{gm}{cm^3} \frac{4\pi}{3} (6.4 \times 10^8)^3 cm^3 = 5.5 \times 10^{27} gm \qquad (20)$$

**The standard gravitational parameter**
As previously told, the standard gravitational parameter μ of a celestial body is the product of the gravitational constant *G* and the mass *M* of the body. The units of the standard gravitational parameter are $cm^3 sec^{-2}$ [18]. The gravitational parameter *GM* is depending on the body concerned, therefore it may also be called the geocentric (Earth) or heliocentric (Sun) gravitational constant, among other names. This quantity gives a convenient simplification of some formulation of the gravitation. Let us remark that for the Earth and the Sun, the value of the product *GM* is known with more accuracy than each factor independently.
Let us assume that in a gravitational interaction between two masses, *m* and *M*, we have m<<M. Then there is a large body *M* at the centre of the orbiting smaller body *m*. Therefore, μ=*GM* is the standard gravitational parameter of the larger body.
For circular orbits around *M*:

$$\mu = MG = r v^2 = r^3 \omega^2 = 4\pi^2 r^3 T^{-2} \qquad (21)$$

Here *r* is the orbit radius, *v* is the orbital speed, *ω* is the angular speed, and *T* is the orbital period. We can generalize for elliptic orbits as:

$$\mu = 4\pi^2 a^3 T^{-2} \rightarrow T^2 \propto a^3 \qquad (22)$$

*a* is the semi-major axis. Therefore, we can tell that the square of the orbital period of a planet is directly proportional to the cube of the semi-major axis of its orbit. This is the third Kepler's law, published by Kepler in 1619.
The large part of the ancient astronomers believed that it was the Sun to move and had a picture of the solar system as that given by Pliny the Elder [6], in the chapter on the motion of planets.
It is as follow. "It is certain that the star called Saturn is the highest, and therefore appears the smallest, that he passes through the largest circuit, and that he is at least thirty years in completing it. … the orbit of Jupiter is much lower, and is carried round in twelve years. The next star, Mars, … from its nearness to the sun is carried round in little less than two years. … The path of the Sun consists of 360 degrees; but, in order that the shadow may return to the same point of the dial, we are obliged to add, in each year, five days and the fourth part of a day. On this account an intercalary day is given to every fifth year, that the period of the seasons may agree with that of the Sun. Below the Sun revolves the great star called Venus, wandering with an alternate motion, … It completes the circuit of the Zodiac in 348 days, never receding from the sun more than 46 degrees, … …(Mercury) is carried in a lower orbit, and moves in a course which is quicker by nine days, shining sometimes before the rising of the sun, and at other times after its setting, but never going farther from it than 23 degrees."
The Pliny's description of the orbital period of Saturn, Jupiter and Mars is good. For Venus and Mercury quite wrong. In fact, the orbital period of Saturn is 29.46 years, that of Jupiter is of 11.86 years, and of Mars of 1.88 years. Venus and Mercury have a orbital period of 0.62 and 0.25 years respectively [19].
Let us suppose that we are able to determine the distance Sun-Earth; knowing the periods of the other planes, we can calculate their distances, applying the Kepler's law. Moreover, having the constant $G$, we can have the mass of the Sun.

**The distance to the Sun**
Aristarchus of Samos lived in the 3rd century BCE. In opposition to the common belief of the other astronomers, he told that the Earth revolves around the Sun, besides rotating on its axis. The original Aristarchus' work on the motion of Earth had not survived, but his ideas are known from the discussion on them by Archimedes, in his treatise "The Sand-Reckoner". Archimedes is telling that Aristarchus proposed that the Universe was vastly larger than was commonly believed. Archimedes is also reporting that Aristarchus measured an apparent size of the Sun of ½°, which is about right (32' or 0.53 degrees) [20].
The only survived work by Aristarchus is "On the Sizes and Distances of the Sun and Moon", where he told how he measured the distance of the Sun. He used the triangle formed by the Earth, Sun and Moon: when we observe a half-moon, this triangle is rectangular one (Fig.7). He assumed that the angular distance α between the Moon and the Sun was 87°. Probably, Aristarchus proposed 87° as a lower limit of this angle [21].

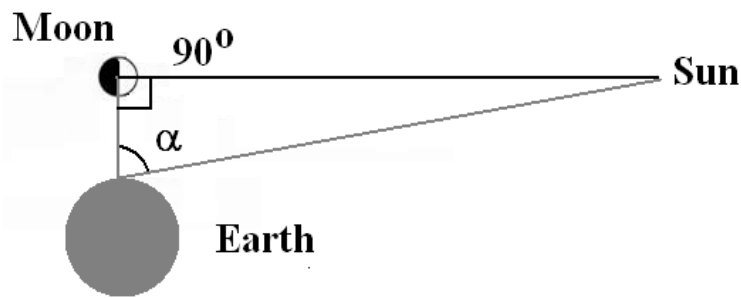

Fig.7

Let us note that an unaided eye can clearly see the line exactly halving the Moon. Aristarchus tried to find the exact moment of the half-moon because at that moment he was sure that the Sun-Moon-Earth angle is exactly 90º. This would allow him to apply a simple triangle geometry: knowing the distance to the Moon – and we have previously discussed that Aristarchus knew this distance very well – he evaluated the distance of the Sun.
But Aristarchus needed to gauge the Sun-Earth-Moon angle. As discussed in [21], when he attempted to measure this other angle, he was amazed finding again 90º. Because no triangle can have two right angles, he reasoned that this angle was imperceptibly close to 90º. But, as told in [21], we are able to distinguish the half-moon because we can appreciate 1' for a disc of the Moon of 30'. For an angle in the space, the equivalence is then "1' is to 30' what 3º are to 90º"' [21]. Therefore, the smallest detectable angle Sun-Earth-Moon is 90º – 3º = 87º, which is exactly what Aristarchus used in the calculations.
Using a good geometry, but a bad angle (87°), Aristarchus obtained a distance to the Sun that was between 18 and 20 times the distance to the Moon. The true value of α is close to 89° 50', and therefore the Sun's distance is about 400 times that of the Moon.
The wrong distance obtained by Aristarchus had as a consequence a wrong estimation of its diameter. Since the Moon and Sun have nearly equal apparent angular sizes, their diameters must be in proportion to their distances from Earth. For Aristarchus the diameter of the Sun was between 18 and 20 times larger than the Moon's diameter. And then the volume of the Sun was 300 times greater than the Earth's volume: this difference of volumes could have inspired the heliocentric model of Aristarchus [22].

If the Moon's distance is known, using the geometry of Fig.7, we have the distance of the Sun and then what we know as the Astronomical Unit (abbreviated as AU, au, a.u., or ua). This is a unit of length equal to approximately the average Earth–Sun distance.

Let go back to our book [1] describing the work of the two astronomers. It asks to evaluate the distance Sun-Earth, using the same geometry as in Fig.7, with an angle α of 89°51'. Let us use the sinus rule for the triangle:

$$D_{Sun} = \frac{D_{Moon}}{\sin(180^o - 90^o - 89^o 51')} = \frac{D_{Moon}}{\frac{9}{60}\frac{2\pi}{360}} = 382 \times 3.84 \times 10^{10} \, cm = 1.46 \times 10^{13} \, cm \quad (23)$$

where $D_{Sun}$ is the distance Sun-Earth and $D_{Moon}$ is the distance Moon-Earth (from Eq.10).

The last problem [1] the two amateur astronomers are facing is the determination of the diameter of the Sun. Again, remembering the observation of Aristarchus, that the apparent diameters of Moon and Sun are equal, we can use the proportion of the distances:

$$\frac{d_{Sun}}{d_{Moon}} = \frac{D_{Sun}}{D_{Moon}} = 382 \quad (24)$$

$$d_{Sun} = 382 \, d_{Moon} = 382 \times 3.48 \times 10^8 \, cm = 13.3 \times 10^{10} \, cm \quad (25)$$

The radius of the Sun is then $6.65 \times 10^{10}$ cm.

**The transit of Venus**

The mean distance from the Earth to the Sun, the astronomical unit, is the unit of measure of the solar system. Knowing this distance and the radius of the Earth we can calculate the solar parallax, as in the left panel of Fig.4, with the Sun instead of the Moon. This is the angle subtended at the Sun by the Earth's mean radius. Using the distance of the Sun given by Eq.23 and the radius of the Earth we have a parallax of approximately 9 seconds of degree. For a distance of $1.49 \times 10^{13}$ cm, we have a parallax of 8.8 seconds of degree. However, we can try to measure the solar parallax, and knowing the mean Earth radius, calculate the astronomical unit [23].

We have seen that the procedure used by Aristarchus to measure the distance to the Sun was good but biased by the observational errors. This procedure based on the geometric principles of parallax last for two thousands of years, until Edmond Halley in 1716 proposed to observe the transit of Venus [23]. The use of Venus transits gave an estimate of $1.53 \times 10^{13}$ cm, 2.6% above the currently accepted value, that of $1.49 \times 10^{13}$ cm [24]. More recently, in 1910, the parallax was measured using the asteroid Eros that passed much closer to Earth than Venus [25].

A transit of Venus happens when this planet passes directly between the Sun and Earth, appearing as a small black disk moving across the Sun bright disk. The duration of such transits is usually measured in hours.

| Venus Transit | 8.842, 8.88 seconds of degree |
| Eros Transit | 8.872, 8.807, 8.79 seconds of degree |
| Venus Radar | 8.796 seconds of degree |
| Doppler Effect | 8.807 seconds of degree |

Values of the solar parallax, from Ref.11.

The transits of Venus are predictable but rare astronomical phenomena. In this century there are two transits: the first was on 8 June 2004 and the second will be on 6 June 2012 [26]. Of course, recent methods based on radar signals and spacecraft telemetry gave the precise value of solar parallax, but the Venus transit of June 2004 provided the opportunity to repeat with modern instruments the measure proposed by Halley.

For the June 2012 transit of Venus, the NASA is inviting the amateur astronomers to participate in education programs designed to inform the public "about this wondrous and historic event" [27]. Therefore, this transit can be a great education event on the astronomy of the solar system. A good opportunity for the two astronomers at Berkeley.